\documentclass[final,5p,twocolumn]{elsarticle}

\usepackage{geometry}
\usepackage{amssymb,amsmath}
\usepackage{epstopdf}
\usepackage{xcolor}
\usepackage{caption}
\usepackage{subcaption}
\usepackage[ruled,vlined]{algorithm2e}
\usepackage{natbib}
\usepackage{bibentry}
\usepackage[hidelinks]{hyperref}

\journal{Finite Elements in Analysis and Design}

\usepackage{tikz}
\usetikzlibrary{arrows.meta}
\tikzset{%
  >={Latex[width=2mm,length=2mm]},
            base/.style = {rectangle, rounded corners, draw=black,
                           minimum width=3cm, minimum height=0.75cm,
                           text centered, font=\sffamily\scriptsize},
}

\makeatletter
\def\ps@pprintTitle{%
 \let\@oddhead\@empty
 \let\@evenhead\@empty
 \def\@oddfoot{}
 \let\@evenfoot\@oddfoot}
\makeatother

\usepackage{charter}

\begin{document}

%\date{}

\begin{frontmatter}



\title{Node-to-node contact-friction problems using run-time parameter updates on a conventional force-deformation finite element$^*$}

% some alternate titles
% Contact-friction problems using parameter updates on zero-length finite elements
%Contact-friction problems using run-time parameter updates on two-noded finite elements
%Contact-friction problems using run-time parameter updates on existing one-dimensional force-deformation elements
%Contact-friction problems using run-time parameter updates on existing one-dimensional elastic or elasto-plastic elements


\author[inst1]{Asifur Rahman}

\affiliation[inst1]{organization={Civil Engineering},%Department and Organization
            addressline={Stony Brook University}, 
            city={Stony Brook},
            postcode={11794}, 
            state={NY},
            country={USA}}

\author[inst2]{Kevin R. Mackie}

\affiliation[inst2]{organization={Civil, Env., and Construction Engineering},%Department and Organization
            addressline={University of Central Florida}, 
            city={Orlando},
            postcode={32816-2450}, 
            state={FL},
            country={USA}}

\begin{abstract}
%% Text of abstract (is there a word limit we need to be aware of?)
A novel implementation of the traditional node-to-node Coulomb contact-friction problem is presented that utilizes run-time parameter updates on conventional elasto-plastic elements. The two-noded elements are defined by an independent uniaxial force-deformation (or constitutive) relation in each degree of freedom. The location of the two nodes may or may not be coincident. A parameter is a pointer to a value (nodal geometry, element property, material property, etc.) in the finite element domain that can be controlled by the user. Parameters that control the frictional interface normal and tangential responses are updated based on contact detection, and eliminate the need for adding new source code to the finite element library of a given software. The run-time algorithm for updating both an elastic and elasto-plastic force-deformation element to achieve a penalty-based contact-friction model is presented. Static and dynamic cases were investigated for a two-noded unit process and compared with the results obtained from closed-form solutions. Multiple interface elements were then used for the sliding, tipping, and rocking responses of a rigid block on rigid foundation. Finally, several case studies were investigated, and the findings were compared with the experimental results or solutions from the literature. The proposed friction-contact implementation can be deployed in larger finite element models, and parameter updates facilitate implementation of a wider array of friction models by changing only the constitutive model.
\end{abstract}


% \begin{highlights}

% \item Innovative idea to solve complex contact-friction problems using simple elastic (or elasto-plastic) materials and spring elements. 

% \item Practical static and dynamic interface problems solved readily without any additional source code changes.

% \item Several static and dynamic cases were investigated and compared with analytical or published results.

% \item The implementation is extensible to wider range of friction models by only changing the constitutive model.

% \item User-driven, run-time algorithm is portable in the future to a wide array of structural and geotechnical nonlinear finite element problems.

% \end{highlights}

%Idea of user-driven algorithms for complex finite element solutions has potential to transform black-box and commercially-driven element/material/physics software. Within the context of the application to structural contact-friction problems specific to this paper, 

% three components to paper
% parameters - exist already
% node-to-node contact - exist already
% novelty lies in solving problems of nonlinear mechanics using run-time model updating


\end{frontmatter}

\makeatletter{\renewcommand*{\@makefnmark}{}
\footnotetext{\textbf{$^*$This is a preprint and may be downloaded for personal use only. Any other use requires prior permission of the author and Elsevier.} Published in \textit{Finite Elements in Analysis and Design} Volume 218, 1 June 2023, 103918, and may be found at: \url{https://doi.org/10.1016/j.finel.2023.103918}}\makeatother}

%\linenumbers


% Introduction, literature, objective statement
% Introduction

\section{Introduction}
% introduction to parameters
Parameters are a method of associating an object with some quantity of interest in the finite element domain (typically), as first described in \citet{scott_parameters}. These quantities can refer to domain components such as nodal geometry, element properties, material properties, or load patterns. They may also refer to output quantities such as nodal displacements, basic element forces, etc. However, there is no limitation on the association, for example, parameters may refer to random variable distribution parameters, design variables, material stages, and properties of the integrator or algorithm. Parameterization may be viewed as a superset of user-defined subroutines in some existing software, whereby users have access to data in the finite element domain. 

% brief review of sparse parameter literature
Typically, parameters have been used pedagogically for conducting (sequential) probabilistic simulations, e.g., \citep{simulation_FEA2}. The concept of updating during the analysis was performed by \citet{thermal_parameters} for reliability analysis for fire loads. Model updating during the analysis has also been implemented by developing new source code in OpenSees for hybrid simulation \citep{model_updating2} and machine learning estimates to match experiments \citep{model_updating1}. However, here it is proposed to use parameters for run-time model updating (user can update parameters during the analysis) without any new source code to solve contact-friction applications. 

% now get into contact-friction as application of parameters
Coulomb friction \citep{coulomb1785theory}, static friction, and viscous friction; these are the three major components of classical friction models \citep{olsson1998friction}. The difference between sliding and rolling friction was demonstrated by Dahl \citep{dahl1968solid, dahl1975solid, dahl1977measurement}. \citet{sorine1993system} showed that friction also depends on the distance traveled after zero velocity crossing rather than solely on velocity. The LuGre model \citep{de1995new} for friction captures experimental frictional behavior like the Stribeck effect, hysteresis, spring-like characteristics for stiction, and varying break-away force. The Leuven model \citep{lampaert2002modification} modified the LuGre model by considering pre-sliding hysteresis, which was further modified by the Maxwell slip model \citep{al2005generalized}.

Large deformation contact problems have mostly been treated by penalty approximations or `trial-and-error' methods \citep{de1991new}. The research of \citet{alart1991mixed} and \citet{simo1992augmented} comprise of applying Newton's method to the saddle-point equations of the augmented Lagrangian formulation. \citet{de1991new} proposed implicit standard material theory that leads to development of another augmented Lagrangian formulation, where the frictional contact problem is treated in a reduced system. Other than the Lagrange multiplier and penalty methods; the  mathematical programming \citep{hung1980frictionless, hongwu1998combined, li2007contact} and complementary methods \citep{kim1996dynamic, zhang2005two} were also developed.

Computational contact problem started with \citet{signorini1933sopra, signorini1959questioni}, who formulated the general problem of the equilibrium of a linear elastic body in frictionless contact. Since the software NASTRAN in late the 1960s, finite element approaches have been widely used in solving computational mechanics problems \citep{yastrebov2013numerical}. The different methods described earlier were also translated into finite element models \citep{shin2008domain}. In the finite element approach, three types of contact discretization techniques are used \citep{yastrebov2013numerical}. These are node-to-node (NTN) \citep{francavilla1975note, taylor1991patch}, node-to-surface (NTS) \citep{hughes1977finite, hallquist1985sliding, zavarise2009modified} and surface-to-surface (STS) \citep{simo1985perturbed, zavarise1998segment, wohlmuth2001iterative}.

% objective and plan
The objective of this paper is to implement a node-to-node Coulomb dry friction interface using only existing conventional finite elements. No additional materials, elements, or formulations are coded explicitly in the source code or through user-defined subroutines of the finite element software. Rather, it is demonstrated in this paper that contact-friction problems can be solved using run-time updates to parameters that describe properties of common uniaxial materials (elastic or elasto-plastic) and elements (two-noded elements with unaxial force-deformation relation in each degree of freedom). Such a user-defined run-time approach is implemented in OpenSees \cite{mckenna2010nonlinear} as a single unit process and demonstrated under static and dynamic loads compared with known solutions. Thereafter the unit process is replicated to finite length interfaces in a wide array of problems ranging from rigid block on rigid foundation, to bolted joints, and flexible cantilevers in bending.

% Proposed modeling technique and OpenSees implementation, schematic
% Modeling
% status: done

\section{Description of The Proposed Model}
\begin{figure*}[!htb]
    \centering
    \begin{subfigure}[b]{.20\textwidth}
        \centering\includegraphics[scale=0.9]{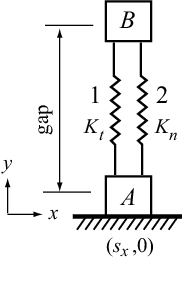}
        \caption{}
        \label{fig:conceptual_1}
    \end{subfigure}~
    \begin{subfigure}[b]{.22\textwidth}
        \centering\includegraphics[scale=0.9]{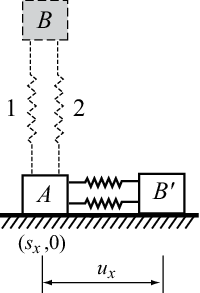}
        \caption{}
        \label{fig:conceptual_2}
    \end{subfigure}~
    \begin{subfigure}[b]{.30\textwidth}
        \centering\includegraphics[scale=0.9]{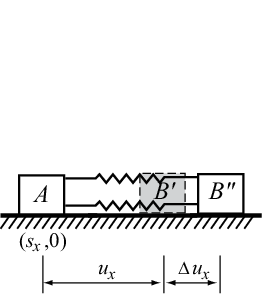}
        \caption{}
        \label{fig:conceptual_3}
    \end{subfigure}~
    \begin{subfigure}[b]{.25\textwidth}
        \centering\includegraphics[scale=0.9]{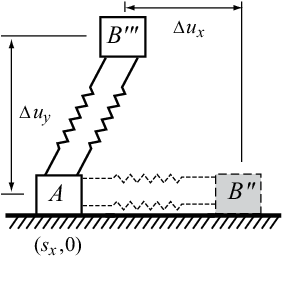}
        \caption{}
        \label{fig:conceptual_4}
    \end{subfigure}~
    
    \caption{Conceptual view of the model using two nodes}
    \label{fig:conceptual}
\end{figure*}

The proposed model is illustrated based on the response of a single unit process connecting two nodes in the finite element model in Figure~\ref{fig:conceptual}. This unit configuration can be replicated as many times as necessary to generate more complex contact-friction interfaces. This section presents the algorithm for parameter updating and the implementation in OpenSees. The implementation is based on parameterization of existing elements and constitutive models; therefore, does not require any additional source code. Parameter updates occur entirely through the interpreter used to make the model input file. Therefore, the unit behavior can be integrated readily into existing models users have already developed. 
% parameterization may be viewed as a generalization of user subroutine whereby behavior of the user subroutine can be controlled at runtime. 

% conceptual view figure

The illustrated behavior assumes node $A$ is fixed and the contact plane is the $x$-axis; however, the configuration of two nodes can be placed anywhere in an existing model, not necessarily attached to fixity. Therefore, the figure and algorithm presented here assume the model is formulated in the local coordinate system shown in Figure~\ref{fig:conceptual_1}. The model is more easily visualized with an element of finite vertical dimension (labeled with the initial gap shown in the figure); however, nodes $A$ and $B$ may have the same nodal geometry (zero gap). The local axis of the element is defined by the corresponding outward normal vector that describes the contact plane ($y$ axis in Figure~\ref{fig:conceptual}). 

Two elements (labeled 1 and 2) connect nodes $A$ and $B$ and represent the horizontal and vertical resistance, respectively. The properties of the model are the gap and normal stiffness $K_n$ for element 2, and the constant cohesion or adhesion term $c$, coefficient of static friction $\mu_s$, and tangential stiffness $K_t$ for element 1. The numerically large normal and tangential stiffness values are penalties for enforcing displacements where contact and slip occur, respectively, and are consistent with other contact-friction implementations. The scalar cohesion is the simplest correction factor for molecular adhesion or other non-pressure-dependent friction sources at the contact surface. 

The parameters of the model are the modulus $K_t$, yield force $F_{\max}$, and initial displacement $s_x$ of element 1, and the modulus $K_n$ of element 2. The gap in the normal direction can be viewed as either achieved through nodal geometry, or resolved internally with an additional initial displacement parameter for element 2. The initial displacement is needed to track the stick (contact) point in the model that can change during the analysis. The stick points are denoted $s_x$ and $s_y$ for each local axis, respectively. The logic sequence of parameter updates in the model are presented in Algorithm~\ref{algo}. 

The constitutive models for both elements can be linear elastic, with secant updates to $K_n$ and $K_t$ to achieve the desired force state at each time step. However, to prevent overshoot of the maximum mobilized tangential friction force, and therefore the need to use smaller time steps, it is easier numerically to use an elasto-plastic constitutive model in element 1. The constitutive model iterates during each step to achieve the correct state, rather than having overshoot from a linear secant model that updates only every increment. In a similar manner, the compression-only behavior of the normal interface can be implemented with a one-sided elastic-plastic gapping constitutive model in element 2 to allow internal iteration. 
\begin{algorithm}[!htb]
\footnotesize
\SetAlgoLined
\KwIn{Initial values of gap, $K_n$, $K_t$, $c$, $\mu$}
\KwOut{Parameter updates at each time step}
\textbf{initialize} \texttt{inContact = false} for all elements, stick point coordinate $(s_x,s_y)\equiv(0,0)$  for all elements\;
\While{analysis continues}{
	\textbf{get} current nodal displacements $(U_x, U_y)$ for $A$ and $B$\;
    \textbf{set} current local displacement, e.g., $u_x = U_{x,B} - U_{x,A}$\;
    \eIf{element is not in contact}{  
        \textbf{set} \texttt{inContact = false}\;
        \textbf{parameter update} $K_t$ = $F_{\max}$ = 0\;
    }{
        \If{new contact}{
            \texttt{inContact = true}\;
            \textbf{parameter update} $(s_x,s_y) = (u_x,u_y)$\;
            \textbf{parameter update} $K_t$ = $K_{t,initial}$\;
        }
            
        \textbf{set} local trial force $F_{x,\mathrm{trial}} = K_t (u_x -  s_x)$\;
        \textbf{set} local trial force $F_{y,\mathrm{trial}} = K_n (u_y -  s_y)$ or read element normal force\;
        \textbf{parameter update} maximum friction $F_{\mathrm{max}} = c + \mu_s F_y$\;
            
        \If{$F_{x,\mathrm{trial}} > F_{\mathrm{max}}$ and linear elastic constitutive model for element 1 }{
            \textbf{parameter update} $s_x = u_x - F_{\mathrm{max}} / K_{t,initial}$\;
            \textbf{parameter update} $K_t = F_{\mathrm{max}} / u_x$\;
        }
            %\eIf{$F_{\mathrm{curr}} \leq F_{\mathrm{max}}$ }{
            %    set $F = F_{\mathrm{curr}}$
            %}{
            %    set $F = F_{\mathrm{max}}$
            %}
    }
    time stepping integrator advances time using standard global assembly
}
\caption{Parameter updating algorithm for unit process}
\label{algo}
\end{algorithm}
%
% schematic and brief description of states of the model
Several states of the unit process are visualized in Figure~\ref{fig:conceptual}. Contact detection is performed using the relative displacements ($u_x,u_y$) between the two nodes (and any initial displacements), as shown in Figure~\ref{fig:conceptual_2}. 
%After contact detection (Figure~\ref{fig:conceptual_2}), the tangent stiffness in the normal and tangential directions are updated to $K_n$ and $K_t$, respectively. Conversely, upon separation (Figure~\ref{fig:conceptual_4}), the tangent stiffnesses are set to zero again. 
At contact detection (point $B'$), the stick point is updated with the displacement $u_x$. While sliding during contact ($B'$ to $B''$ in Figure~\ref{fig:conceptual_3}), the normal force $F_y$ in element 2 is used to update the maximum friction that can be mobilized in element 1. The trial friction force is $F_x = K_t (u_x-s_x+\Delta u_x)$, but is limited to the maximum achievable force $F_{\max}$. Uplift from the contact plane (point $B'''$ in Figure~\ref{fig:conceptual_4}) can occur with a $u_y = \Delta u_y > 0$, at which time contact is lost and the resisting force and tangent stiffness are updated again to zero. 

The initial $K_n$ needs to be sufficiently large to prevent node $B$ moving past point $A$ (in negative $y$ direction). Whereas $K_t$ approaching infinity would reproduce the step function implied by dry friction models, as long as $F_{\max}/K_t \ll u_{x,\max}$, the initial $K_t$ value should be selected to avoid numerical conditioning issues.

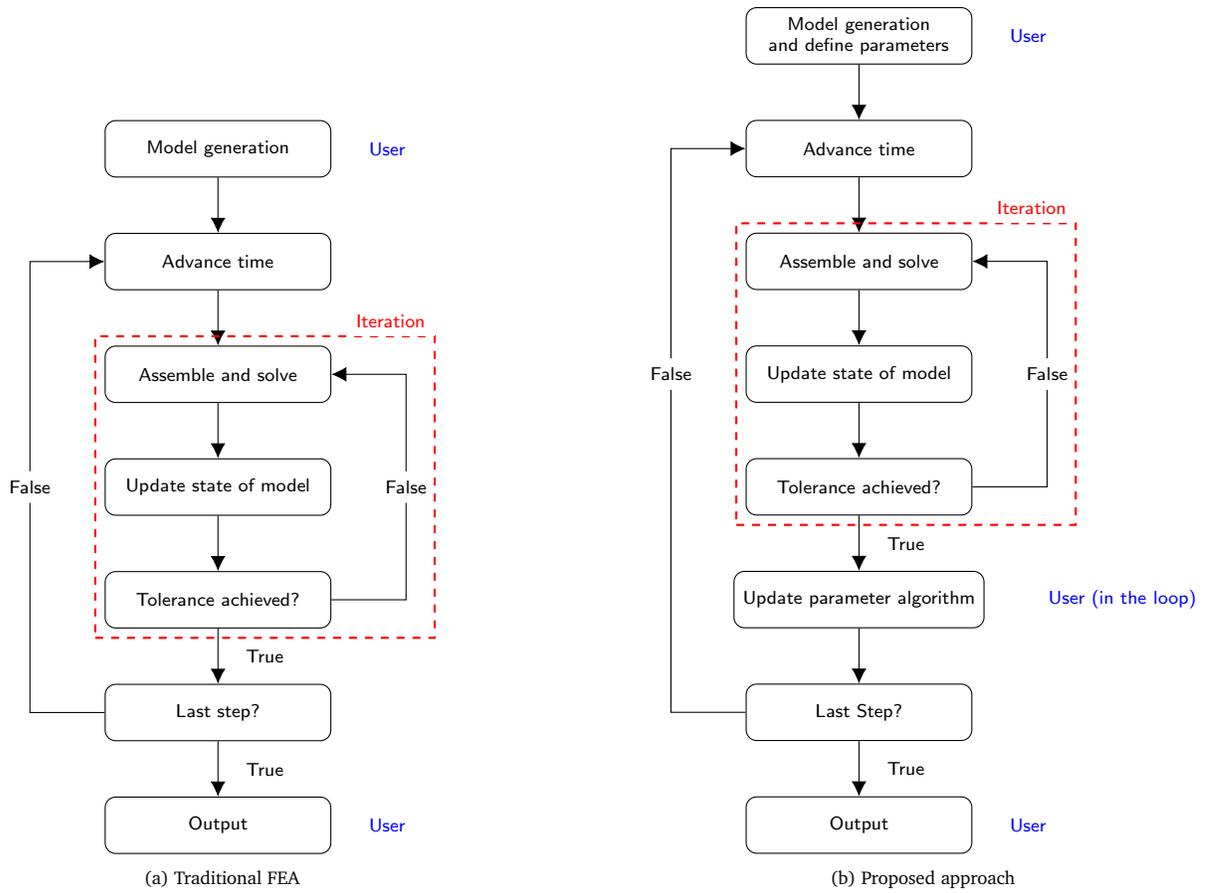
\begin{figure*}[!htb]
\centering
    \begin{subfigure}{.5\textwidth}
        \centering
        \begin{tikzpicture}[node distance=1.5cm,
        every node/.style={fill=white, font=\sffamily\scriptsize}, align=center]
          % Specification of nodes (position, etc.)
          \node (model) [base] {Model generation};
          \node (user1) [right of=model, xshift=0.75cm, text=blue] {User};
          \node (advanceTime) [base, below of=model]  {Advance time};
          \node (assemble) [base, below of=advanceTime] {Assemble and solve};
          \node (update) [base, below of=assemble] {Update state of model};
          \node (tolerance) [base, below of=update] {Tolerance achieved?};
          \node (lastStep) [base, below of=tolerance] {Last step?};
          \node (false2) [right of=update, xshift=1cm] {False};
          \node (output) [base, below of=lastStep] {Output};
          \node (false1) [left of=update, xshift=-1cm] {False};
          \node (user2) [right of=output, xshift=0.75cm, text=blue] {User};
                                                  
          % Specification of lines between nodes specified above with aditional nodes for description 
          \draw[->] (model) -- (advanceTime);
          \draw[->] (advanceTime) --  (assemble);
          \draw[->] (assemble) -- (update);
          \draw[->] (update) -- (tolerance);
          \draw[->] (tolerance) -- (lastStep) node[midway, right=0.25cm] {True};
          \draw[->] (lastStep) -- (output) node[midway, right=0.25cm] {True};
          \draw[->] (lastStep) -| (false1) |- (advanceTime);
          \draw[->] (tolerance) -| (false2) |- (assemble);
        
          \node (test1) [above left] at (assemble.north west) {};
          \node (test2) [above right,xshift=1.25cm] at (assemble.north east) {};
          \node (test3) [below right,xshift=1.25cm] at (tolerance.south east) {};
          \node (test4) [below left] at (tolerance.south west) {};
          \draw[dashed,thick,red] (test1.center) -- (test2.center) -- (test3.center) -- (test4.center) -- cycle;
          \node [text=red,above left] at (test2.center) {Iteration};
          \end{tikzpicture}
        \caption{Traditional FEA}
        \label{fig:flowchart_traditional}
    \end{subfigure}~
    \begin{subfigure}{.5\textwidth}
        \centering
        \begin{tikzpicture}[node distance=1.5cm,
        every node/.style={fill=white, font=\sffamily\scriptsize}, align=center]
          % Specification of nodes (position, etc.)
          \node (model) [base] {Model generation\\and define parameters};
          \node (user1) [right of=model, xshift=0.75cm, text=blue] {User};
          \node (advanceTime) [base, below of=model] {Advance time};
          \node (assemble) [base, below of=advanceTime] {Assemble and solve};
          \node (update) [base, below of=assemble] {Update state of model};
          \node (tolerance) [base, below of=update] {Tolerance achieved?};
          \node (updateParameter) [base, below of=tolerance] {Update parameter algorithm};
          \node (user3) [right of=updateParameter, xshift=2cm, text=blue] {User (in the loop)};
          \node (lastStep) [base, below of=updateParameter] {Last Step?};
          \node (output) [base, below of=lastStep] {Output};
          \node (false2) [right of=update, xshift=1cm] {False};
          \node (user4) [right of=output, xshift=0.75cm, text=blue] {User};
          \node (false1) [left of=update, xshift=-1cm] {False};
                                                  
          % Specification of lines between nodes specified above
          % with aditional nodes for description 
          \draw[->] (model) -- (advanceTime);
          \draw[->] (advanceTime) -- (assemble);
          \draw[->] (assemble) -- (update);
          \draw[->] (update) -- (tolerance);
          \draw[->] (tolerance) -- (updateParameter) node[midway, right=0.25cm] {True};
          \draw[->] (updateParameter) -- (lastStep);
          \draw[->] (lastStep) -- (output) node[midway, right=0.25cm] {True};
          \draw[->] (lastStep) -| (false1) |- (advanceTime);
          \draw[->] (tolerance) -| (false2) |- (assemble);

          \node (test1) [above left] at (assemble.north west) {};
          \node (test2) [above right,xshift=1.25cm] at (assemble.north east) {};
          \node (test3) [below right,xshift=1.25cm] at (tolerance.south east) {};
          \node (test4) [below left] at (tolerance.south west) {};
          \draw[dashed,thick,red] (test1.center) -- (test2.center) -- (test3.center) -- (test4.center) -- cycle;
          \node [text=red,above left] at (test2.center) {Iteration};
          \end{tikzpicture}
        \caption{Proposed approach}
        \label{fig:flowchart_proposed}
    \end{subfigure}
\caption{Flowchart of information in parameter updating of unit process with respect to traditional finite element analysis}
\label{fig:flowchart}
\end{figure*}

% integration of unit process into FEA
The integration of the parameter updating algorithm for the unit process into the finite element analysis solution procedure is illustrated schematically in Figure \ref{fig:flowchart}. The time/load stepping integrator as well as the iterative solution of the nonlinear system of equations are still retained. For portions of the model that are not parameterized, state update would still occur during the iterations. 
%
% illustrate schematically the parameter updating process with respect to traditional finite elements

 % extension to 3D model
The model, validation cases, and case studies presented here are two-dimensional implementations. However, the method is also applicable to three-dimensional contact friction problems using the same local coordinates in Figure~\ref{fig:conceptual} and Algorithm~\ref{algo}. The additional quantities for displacement and force in the local $z$-axis are necessary and correspond to an additional element 3. Methodologically, an additional step is necessary to resolve the maximum frictional force $F_{\max}$ into $F_x$ and $F_z$. This guarantees $F_{\max}$ is the maximum achievable resistance regardless of the in-plane direction of load.

% Validation using two nodes only, nonlinear static and nonlinear dynamic cases compared to analytical solutions
% Validation

\section{Validation}
Several load cases were investigated to validate the run-time parameter updating process using the single unit process shown in Figure~\ref{fig:conceptual}. They are separated into static and dynamic loads, with the analytically correct solution generated for each comparison. Both the elastic secant and elasto-plastic constitutive model approaches are presented for comparison. The numerical solutions were generated using the software framework OpenSees. The OpenSees element denoted \emph{zeroLength} was used for both element 1 and element 2. The elastic secant approach used the \emph{ElasticMaterial} constitutive model, whereas the elasto-plastic approach used the \emph{ElasticPP} and \emph{ElasticPPGap} constitutive models for the tangential and normal directions, respectively. 

% ----------------------------------------------------------------------------------------------------------------------------------------
\subsection{Nonlinear Static Cases}
\begin{figure*}[!htb]
\centering
\begin{subfigure}{.3\textwidth}
\centering\includegraphics[width=0.75\textwidth]{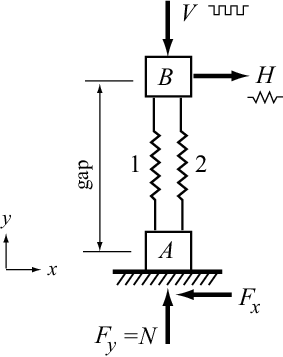}
\caption{Schematic}
\label{fig:two_node_static_schematic}
\end{subfigure}~
\begin{subfigure}{.5\textwidth}
% note that former two_node_static_1 and two_node_static_2 need to get combined in single plot that is called below
\centering\includegraphics[width=\textwidth]{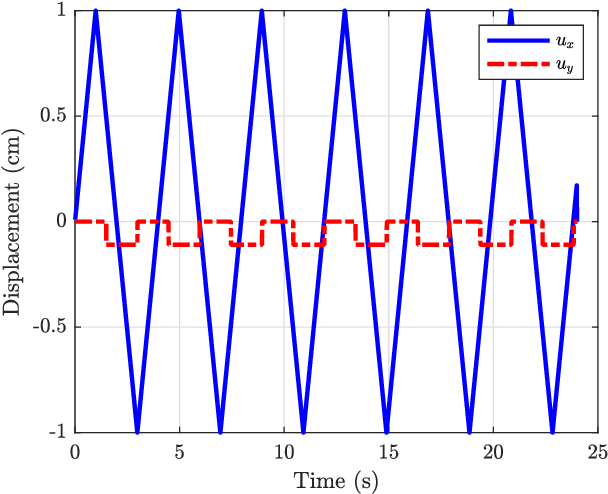}
\caption{Applied displacements}
\label{fig:two_node_static_load}
\end{subfigure}
\caption{Two-noded static validation case model and loading}
\label{fig:two_node_static}
\end{figure*}

\begin{figure*}[!htb]
\centering
\begin{subfigure}{.5\textwidth}
\centering\includegraphics[width=\textwidth]{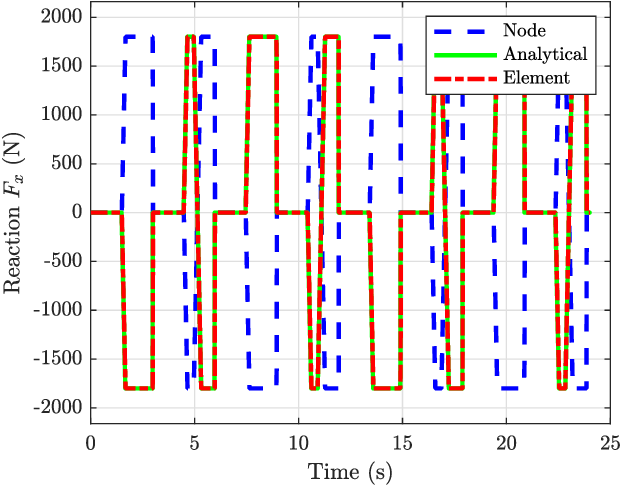}
\caption{Horizontal reaction}
\label{fig:two_node_static_fx}
\end{subfigure}~
\begin{subfigure}{.5\textwidth}
\centering\includegraphics[width=\textwidth]{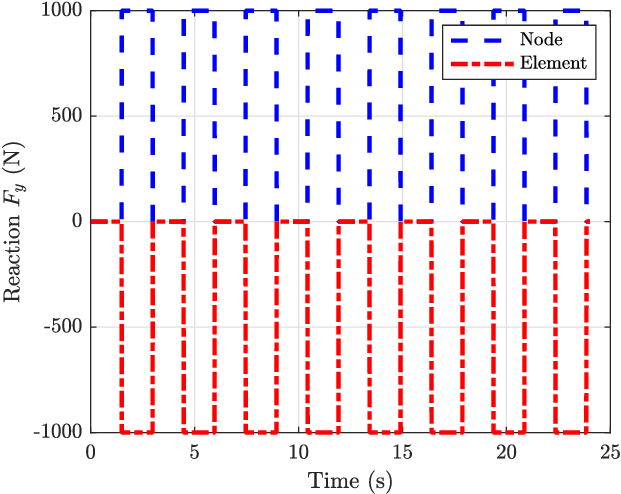}
\caption{Vertical reaction}
\label{fig:two_node_static_fy}
\end{subfigure}
 
\begin{subfigure}{.5\textwidth}
\centering\includegraphics[width=\textwidth]{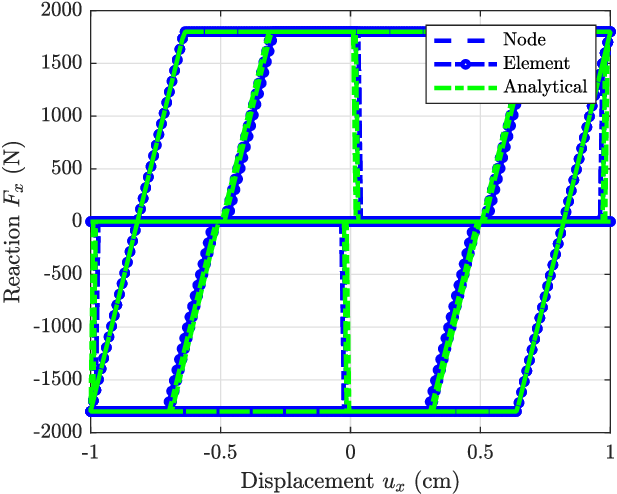}
\caption{Element horizontal hysteresis}
\label{fig:two_node_static_fxux}
\end{subfigure}~
\begin{subfigure}{.5\textwidth}
\centering\includegraphics[width=\textwidth]{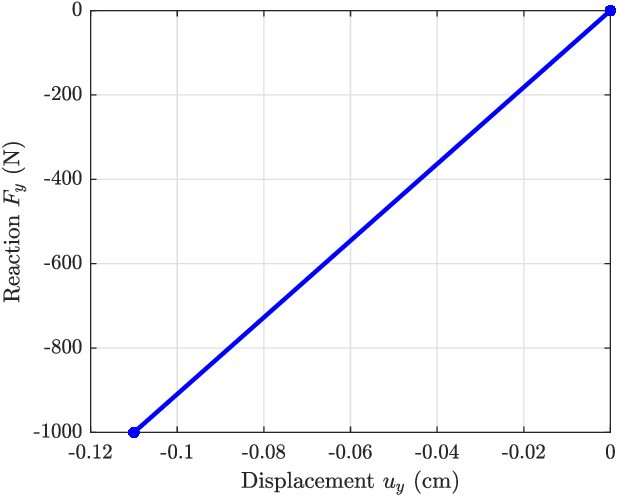}
\caption{Element vertical hysteresis}
\label{fig:two_node_static_fyuy}
\end{subfigure}

\caption{Two-noded static validation case results using elasto-plastic modeling approach}
\label{fig:two_node_static_results}
\end{figure*}

The two-noded static validation case model is shown in Figure~\ref{fig:two_node_static_schematic}. Nodes $A$ and $B$ were offset with a vertical gap of 0.1 cm. Node $B$ was subjected to horizontal periodic (period of 4 s) triangular displacements and vertical periodic (period of 3 s) pulse displacements, as shown in Figure~\ref{fig:two_node_static_load}. These two displacements were set out of phase purposefully to demonstrate that the unit process can correctly update the stick point during the analysis. The amplitude of the horizontal and vertical displacements were 1.0 cm and 0.11 cm, respectively. A time step of 0.01 s was used (static analysis). The parameters of the numerical model were $K_n$, $K_t$, $\mu_s$, and $c$ of 1.0$e^5$ N/cm, 1.0$e^4$ N/cm, 0.3, and 1.5$e^3$ N, respectively. 
%
% two-noded static case
Figures \ref{fig:two_node_static_fx} and \ref{fig:two_node_static_fxux} show the reaction time histories $F_x$ and $F_y$, respectively, from numerical modeling with the elasto-plastic approaches match the analytical results. The nodal reactions are the opposite signs to the element resisting forces. The correct stick point behavior is obtained as contact is made and lost repeatedly during the horizontal displacement history. The corresponding hysteresis plots are shown in Figures~\ref{fig:two_node_static_fxux} and ~\ref{fig:two_node_static_fyuy} for the horizontal and vertical directions, respectively. The effect of the finite $K_t$ and $K_n$ values selected in the validation case study manifests itself as the stiffness shown in the hysteresis plots. As the penalty factors are increased, the frictional hysteretic behavior tends toward rigid plastic as would be expected for dry friction. 

% two_noded static results

\begin{figure*}[!htb]
\centering
\begin{subfigure}{.5\textwidth}
\centering\includegraphics[width=\textwidth]{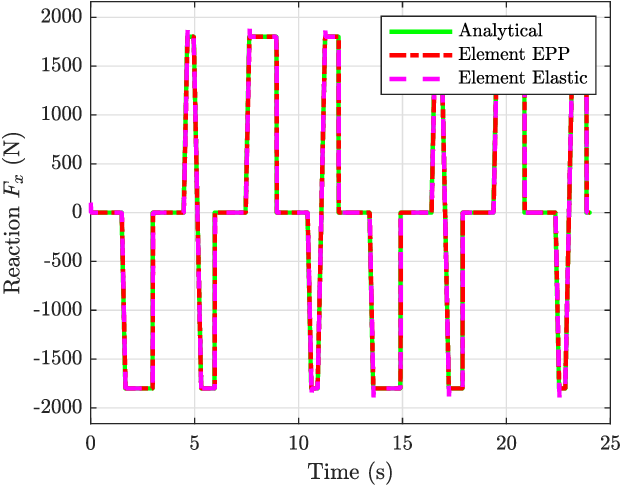}
\caption{Horizontal reaction}
\label{fig:two_node_static_fx_compare}
\end{subfigure}~
\begin{subfigure}{.5\textwidth}
\centering\includegraphics[width=\textwidth]{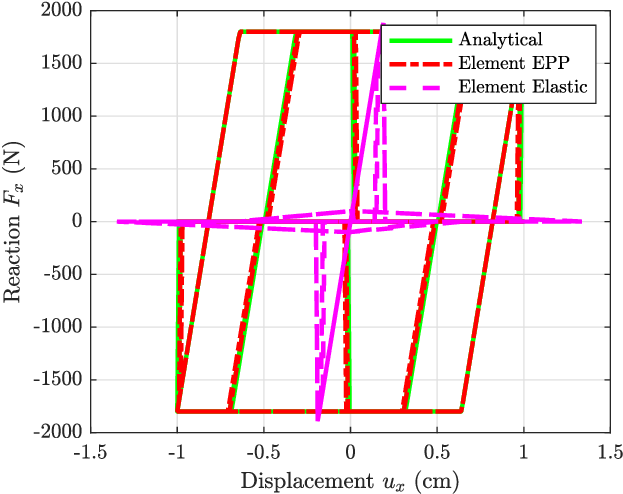}
\caption{Element horizontal hysteresis}
\label{fig:two_node_static_fxux_compare}
\end{subfigure}
\caption{Two-noded static validation case results comparing elasto-plastic with elastic modeling approaches}
\label{fig:two_node_static_compare}
\end{figure*}

To demonstrate the difference in the two implementations (elasto-plastic and elastic), the frictional responses are compared in Figure~\ref{fig:two_node_static_compare}. The friction reaction histories in Figure~\ref{fig:two_node_static_fx_compare} demonstrate both approaches follow the analytical results; however, the secant approach leads to a force overshoot when sliding initiates. The size of the force overshoot depends on the time step adopted in the analysis and the penalty $K_t$. The overshoot is caused because the parameter updates occur at the increment level (time step), not at the iteration level as is does within the elasto-plastic constitutive models. As the loads do not reverse rapidly in the nonlinear static cases, the overshoot is corrected in the subsequent analysis step. The hysteretic comparison in Figure~\ref{fig:two_node_static_fxux_compare} provides more insight into the stick point in Algorithm~\ref{algo}. The elastic approach develops larger element deformations even though compatibility at the nodes (nodal displacements) are the same with both modeling approaches. 

% two_noded static results comparing elasto-plastic with elastic modeling approaches

% ----------------------------------------------------------------------------------------------------------------------------------------
\subsection{Nonlinear Dynamic Cases}
\begin{figure*}[!htb]
\centering
\begin{subfigure}{.3\textwidth}
\centering\includegraphics[width=0.75\textwidth]{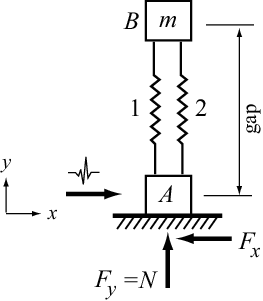}
\caption{Schematic}
\label{fig:two_node_dynamic_schematic}
\end{subfigure}~
\begin{subfigure}{.5\textwidth}
% note that former two_node_dynamic_1 and two_node_dynamic_2 need to get combined in single plot that is called below
\centering\includegraphics[width=\textwidth]{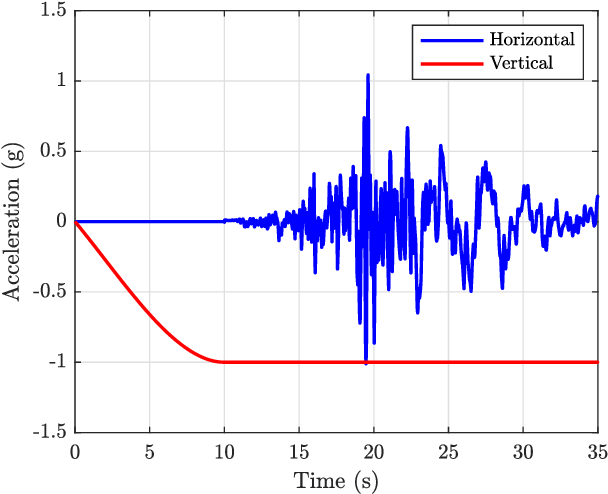}
\caption{Applied accelerations}
\label{fig:two_node_dynamic_load}
\end{subfigure}
\caption{Two-noded dynamic validation case model and loading}
\label{fig:two_node_dynamic}
\end{figure*}

The two-noded dynamic validation case model is shown in Figure~\ref{fig:two_node_dynamic_schematic}. Nodes $A$ and $B$ were placed at the same location (no initial gap). Node $B$ was assigned a mass of 4.0 kg and acceleration time histories were applied to the model, as shown in Figure~\ref{fig:two_node_dynamic_load}. The vertical acceleration was gradually increased to 1 g for gravity load to avoid vibrations in the response, after which the SVL360 (Loma Prieta 1989 Sunnyvale - Colton Avenue station) ground motion was applied horizontally. The ground motion was scaled by a factor of 5 to achieve a similar range of displacements as were observed during the nonlinear static validation cases. A time step of 0.005 s was used (transient analysis). The parameters of the numerical model were $K_n$, $K_t$, $\mu_s$, and $c$ of 1.0$e^5$ N/cm, 1.0$e^6$ N/cm, 0.3, and 1.5$e^3$ N, respectively. 

% I'm tempted to move the SDOF equation of motion here and solve it with ode45. 

% note originally we had perhaps intended to do acceleration histories that contained two parts, but not sure they are necessary now
%- first horizontal only, but with 1g vertical constant acceleration
%- then both horizontal and vertical with different ground motions

% two-noded dynamic case

\begin{figure*}[!htb]
\centering
\begin{subfigure}{.5\textwidth}
\centering\includegraphics[width=\textwidth]{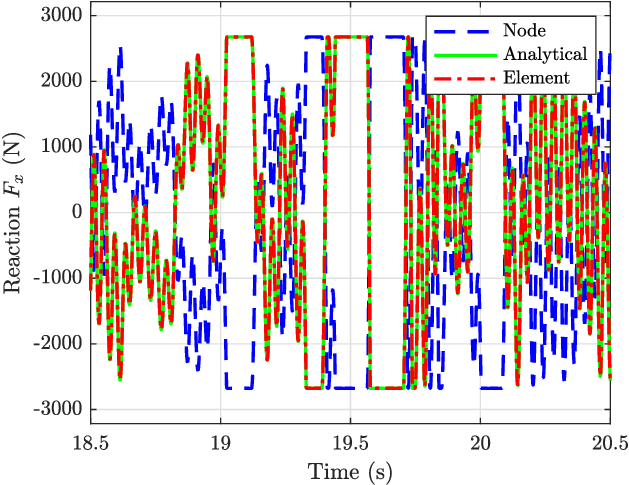}
\caption{Horizontal reaction}
\label{fig:two_node_dynamic_fx}
\end{subfigure}~
\begin{subfigure}{.5\textwidth}
\centering\includegraphics[width=\textwidth]{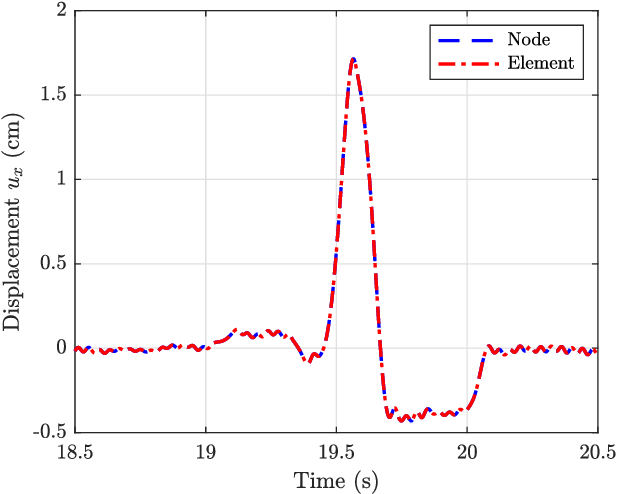}
\caption{Horizontal displacement}
\label{fig:two_node_dynamic_ux}
\end{subfigure}
 
\begin{subfigure}{.5\textwidth}
\centering\includegraphics[width=\textwidth]{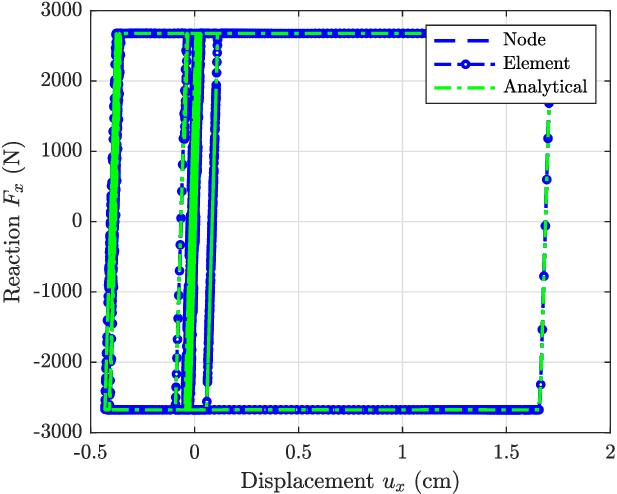}
\caption{Element horizontal hysteresis}
\label{fig:two_node_dynamic_fxux}
\end{subfigure}~
\begin{subfigure}{.5\textwidth}
\centering\includegraphics[width=\textwidth]{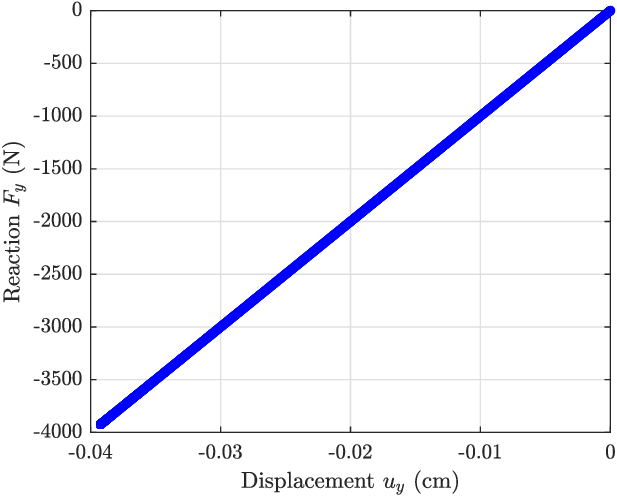}
\caption{Element vertical hysteresis}
\label{fig:two_node_dynamic_fyuy}
\end{subfigure}

\caption{Two-noded dynamic validation case results using elasto-plastic modeling approach}
\label{fig:two_node_dynamic_results}
\end{figure*}

Figures \ref{fig:two_node_dynamic_fx} and \ref{fig:two_node_dynamic_ux} show the reaction $F_x$ and displacement $u_x$ time histories, respectively, from numerical modeling with the elasto-plastic approach (for a subset of the total analysis time to better visualize the response during slipping). The analytical solution was obtained by assuming the nodal displacements from the model were correct and computing the corresponding element forces. 
% We don't have sensitivity yet
%A more rigorous solution of the coupled equations of motion was performed in the sensitivity analysis. 
The nodal reactions are the opposite signs to the element resisting forces, and the element forces agree exactly with the analytical solution. Stick point changes do not occur from loss of contact due to the constant gravity load. The corresponding hysteresis plots are shown in Figures~\ref{fig:two_node_dynamic_fxux} and ~\ref{fig:two_node_dynamic_fyuy} for the horizontal and vertical directions, respectively. 

% may be some value in adding the actual ode45 solution here (that has rigid-plastic excursions) and labeling the Analytical one as something else here

% two_noded dynamic results

\begin{figure*}[!htb]
\centering
\begin{subfigure}{.5\textwidth}
\centering\includegraphics[width=\textwidth]{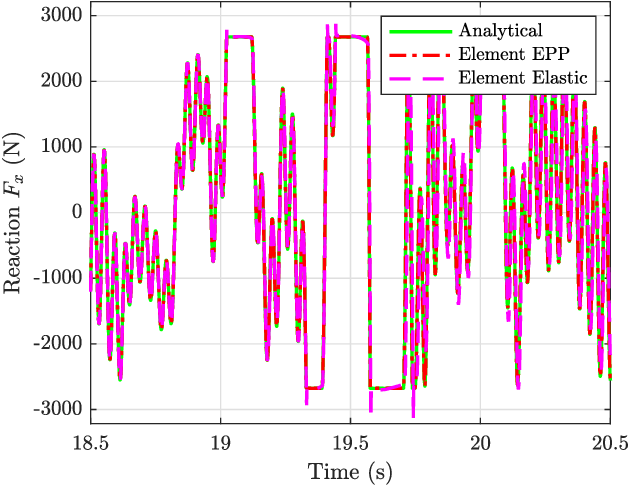}
\caption{Horizontal reaction}
\label{fig:two_node_dynamic_fx_compare}
\end{subfigure}~
\begin{subfigure}{.5\textwidth}
\centering\includegraphics[width=\textwidth]{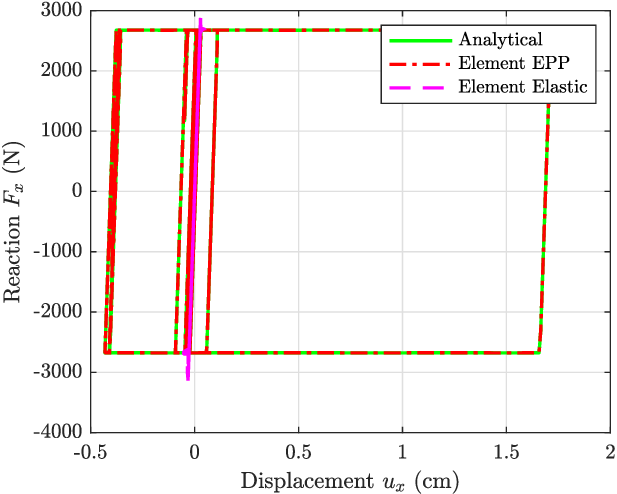}
\caption{Horizontal hysteresis}
\label{fig:two_node_dynamic_fxux_compare}
\end{subfigure}

\caption{Two-noded dynamic validation case results comparing elasto-plastic with elastic modeling approaches}
\label{fig:two_node_dynamic_compare}
\end{figure*}

To demonstrate the difference in the two implementations (elasto-plastic and elastic), the frictional responses are compared in Figure~\ref{fig:two_node_dynamic_compare}. As with the static load cases, the friction reaction histories in Figure~\ref{fig:two_node_dynamic_fx_compare} all show agreement. However, the secant approach leads to a force overshoot when sliding initiates that depends on the time step adopted in the analysis and the penalty $K_t$. More apparent in the dynamic responses is the delay in returning to the correct solution due to both the varying inertial forces, but also the rapid load reversals that occur during the history. Consequently, the peak displacement is 15\% larger in the analysis presented. The hysteretic comparison in Figure~\ref{fig:two_node_dynamic_fxux_compare} provides more insight into the stick point in Algorithm~\ref{algo}. Unlike the static case, the element displacements were substantially smaller than the nodal displacements, because of the single large reversal in the load pattern (and the corresponding updates to $s_x$).

% two_noded dynamic results comparing elasto-plastic with elastic modeling approaches

%\begin{figure}
%\centering
    %\begin{subfigure}{.5\textwidth}
    %\centering\includegraphics[width=\textwidth]{figures/two_node_sensitivity/two_node_contour.eps}
    %\caption{Element forces in sliding case}
    %\label{fig:block_bottom_1}
    %\end{subfigure}
%\caption{Sliding and tipping static case}
%\label{block_static_bottom}
%\end{figure}

% AR start
% need to return to scripts to correctly compute the ratio with respect to the analytical solution

%sensitivity to Kt, Kn, mu\_s, and c
%- investigate how sensitive the dynamic forces (and perhaps displacements) are to the Kn and Kt values
%- what is the necessary dynamic integrator and step size, dependence of results on step size
%- does it work equally well with explicit vs implicit integrators?
%- perhaps mu\_s and c are not necessary here other than including them in the nonlinear static cases above

% create figures that have normalized displacement on the horizontal axis and each of the parameters would show deviation from unit value
% would require normalization of the parameter values to get a meaningful vertical axis as well
% AR end

% Rectangular rigid block studies
% Block
% status: incomplete
% missing tip and slip, dynamic slipping, and contour plot

\section{Rigid Block Studies}
The single unit process is extended here to a finite length contact interface and studied using several classical sliding, tipping, and rocking cases of a rigid block on a rigid surface. 

% ----------------------------------------------------------------------------------------------------------------------------------------
\subsection{Nonlinear Static Tip and Slip}
An analytically rigid block of dimensions 100 mm and 100 mm was placed on a rigid surface (single-point constraints) aligned in the $x$ direction. A total of 11 of the unit processes from Figure \ref{fig:conceptual} were used with an equal spacing of 10 mm. Two classical load cases were considered for the static analysis, one placed at the base of the block to induce pure lateral demand (sliding), and the second placed at the top of the block to induce tipping. For the static slipping case, the cohesion between the block and the contact surface was set to zero. However, for the tipping case, a large cohesion was used.

\begin{figure*}[!htb]
\centering
\begin{subfigure}{.30\textwidth}
\centering\includegraphics[width=\textwidth]{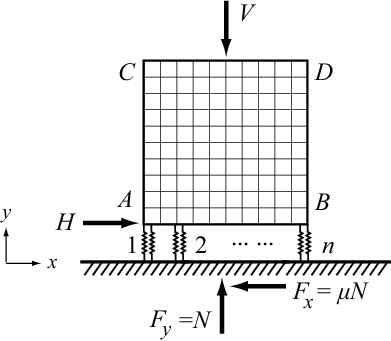}
\caption{Load applied at the base of the block}
\label{fig:block_static_bottom}
\end{subfigure}~
\begin{subfigure}{.30\textwidth}
\centering\includegraphics[width=\textwidth]{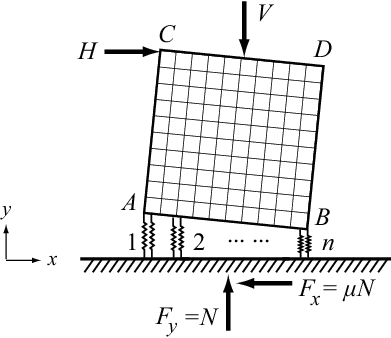}
\caption{Load applied at the top of the block}
\label{fig:block_static_top}
\end{subfigure}~
\begin{subfigure}{.30\textwidth}
\centering\includegraphics[width=\textwidth]{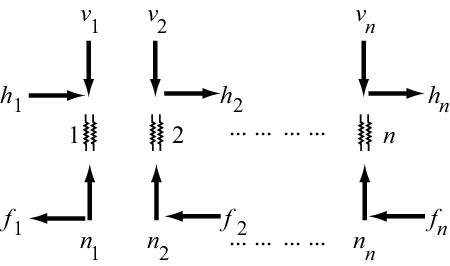}
\caption{Discrete interface elements}
\label{fig:block_fbd}
\end{subfigure}
\caption{Rigid block on rigid foundation static cases}
\label{fig:block_static}
\end{figure*}

Both cases of lateral loading were accompanied by a vertical load $V$ of 1100 N applied at the top-center of the block to simulate the gravity force, and are shown in Figures \ref{fig:block_static_bottom} and \ref{fig:block_static_top}, respectively. The forces acting on the array of interface elements is shown in Figure~\ref{fig:block_fbd}. The parameters of the numerical model were modulus of elasticity $E$, Poisson's ratio $\nu$, thickness $t$, $K_t$, $K_n$, and $\mu$ of 10 GPa, 0, 1 mm, $1e^5$ N/mm, $1e^5$ N/mm, and 0.3 respectively.

% static block figures

Incipient tipping occurs when the resultant of the normal forces at the contact interface reaches the edge of the block. Thereafter, the block experiences rigid-body rotation about this pivot point. Assuming the load (horizontal and vertical) follow the same locations on the block, the kinematic expression for the displacements of point $C$ in the block is shown in Equation \ref{eq:block}.

%- approximately rigid block
%- can pick a 1/1 or 3/1 case for all the examples in this section
%- look at incipient slip and incipient tip cases as compared to statics solutions (including the resultant of the normal distributed reaction force)

\begin{equation}
    u_C = R \left[ \left( \cos\left(\theta^\prime - \theta\right) - \cos\theta^\prime \right) \hat{\mathbf{i}} 
    + 
    \left( \sin\left(\theta^\prime - \theta\right) - \sin\theta^\prime \right) \hat{\mathbf{j}}\right]
    \label{eq:block}
\end{equation}
where $R$ is the length of the diagonal of the block, $\hat{\mathbf{i}}$ and $\hat{\mathbf{j}}$ are unit vectors in $x$ and $y$ directions, $\theta^\prime = \pi/2 + \alpha$, $\theta$ and $\alpha$ are angles as shown in Figure \ref{fig:block_dynamic_rock}.

\begin{figure*}[!htb]
\centering
    \begin{subfigure}{.5\textwidth}
    \centering\includegraphics[width=\textwidth]{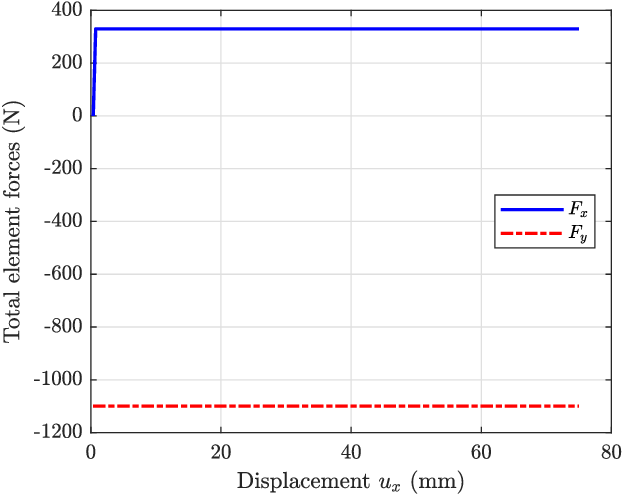}
    \caption{Element forces vs. horizontal displacement in sliding}
    \label{subfig:sliding_tipping_static_1}
    \end{subfigure}~
    \begin{subfigure}{.5\textwidth}
    \centering\includegraphics[width=\textwidth]{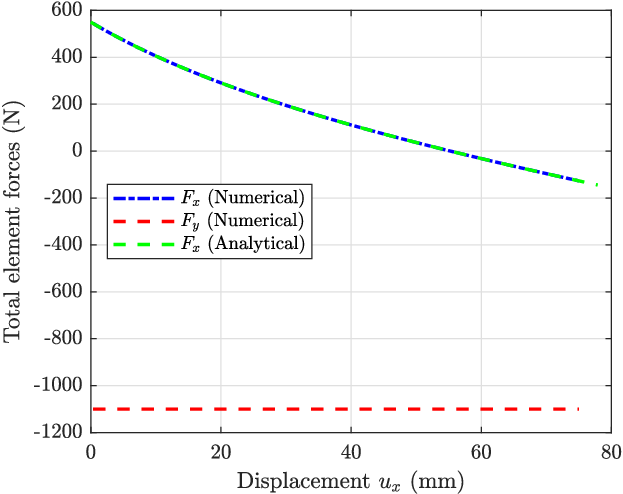}
    \caption{Element forces vs. horizontal displacement in tipping}
    \label{subfig:sliding_tipping_static_2}
    \end{subfigure}

    \begin{subfigure}{.5\textwidth}
    \centering\includegraphics[width=\textwidth]{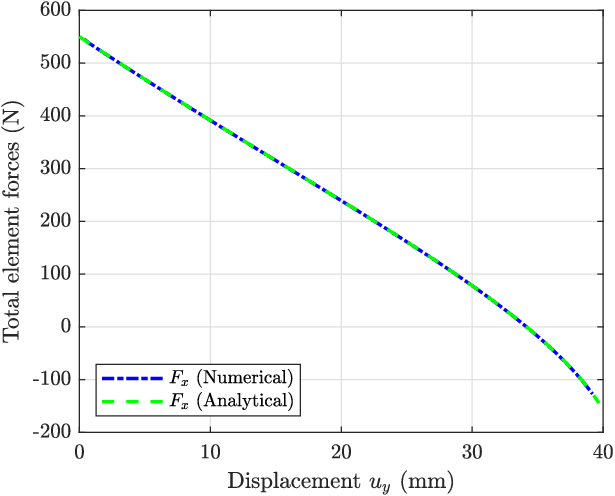}
    \caption{Element forces vs. vertical displacement in tipping}
    \label{subfig:sliding_tipping_static_3}
    \end{subfigure}~
    \begin{subfigure}{.5\textwidth}
    \centering\includegraphics[width=\textwidth]{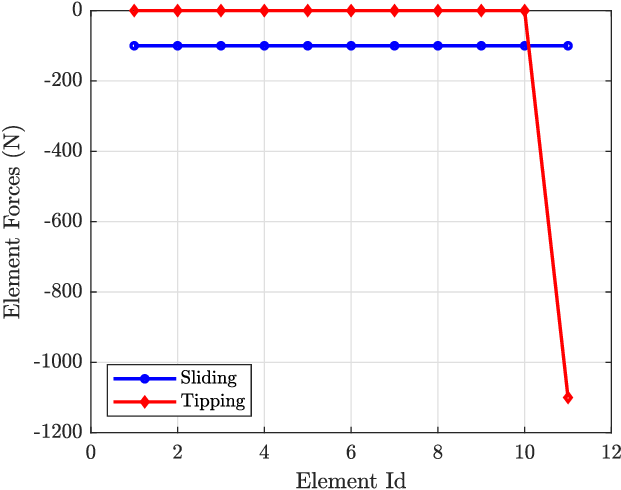}
    \caption{Distribution of element forces}
    \label{subfig:sliding_tipping_static_4}
    \end{subfigure}~
\caption{Sliding and tipping static cases time histories}
\label{fig:sliding_tipping_static}
\end{figure*}

The force vs horizontal deformation results for the static slipping case are presented in Figure \ref{subfig:sliding_tipping_static_1}. The friction force develops linearly proportional to $K_t$, and the block starts slipping once the maximum frictional force that can develop on the contact surface is exceeded. Each contact element develops a normal force of the same magnitude as there is no overturning tendency by the block (horizontal load $H$ applied at base). In both cases of load application, the resultant of the normal forces in the frictional elements is equal to the vertically applied load $V$ on the top of the block.

The force vs horizontal deformation results for the static tipping case are presented in Figure \ref{subfig:sliding_tipping_static_2}. The frictional force develops nonlinearly with the horizontal displacement for tipping case due to equilibrium in the deformed configuration. The relation between the frictional force and vertical displacement of point $C$ shown in Figure \ref{subfig:sliding_tipping_static_3} similarly decreases. As uplift initiates,  the friction elements in tension (separation) have zero normal forces, whereas higher normal forces appear on the compression side as shown in Figure \ref{subfig:sliding_tipping_static_4}. 

% static results

%Figure \ref{subfig:sliding_tipping_static_1} shows the normal force and friction force history for the sliding case. The block starts sliding once it develops the maximum frictional force allowed by the contact surface while the normal forces remain same during sliding. Figure \ref{subfig:sliding_tipping_static_2} and \ref{subfig:sliding_tipping_static_3} show the horizontal and vertical displacement history for the top right corner node $D$ of the block for the tipping case. The final subfigure \ref{subfig:sliding_tipping_static_4} compares the distribution of normal forces in the contact elements for the sliding and tipping cases. In sliding case, all elements share the normal force equally, but in the tipping case, the bottom corner node $B$ takes all the weight of the block as expected.

% ----------------------------------------------------------------------------------------------------------------------------------------
\subsection{Nonlinear Dynamic Slipping}

\begin{figure*}[!htb]
\centering\includegraphics[height=1.8in]{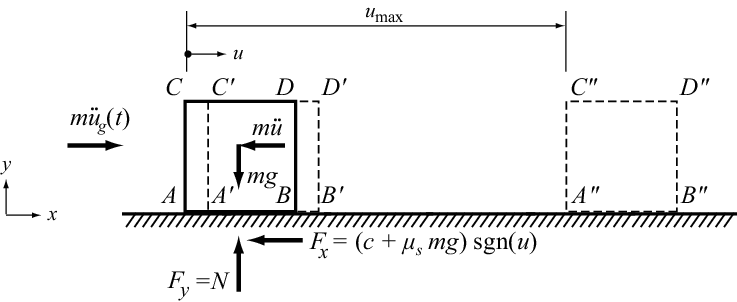}
\caption{Dynamic block sliding schematic}
\label{fig:block_dynamic_slide}
\end{figure*}

The use of multiple unit processes are also able to capture dynamic sliding and tipping response of a block, as shown in Figure \ref{fig:block_dynamic_slide}. The equation of motion for pure sliding (single dynamic degree of freedom) is shown in Equation \ref{eq:sliding}. The block starts moving as soon as the inertia force $m\ddot{u}_g$ exceeds the frictional resistance $c+\mu_s mg$, where, $\ddot{u}_g$ is the ground acceleration, $g$ is the acceleration due to gravity, $\mu$ is the coefficient of friction and $c$ is the cohesion or adhesion between the block and the contact surface. Unlike the two-noded validation case; however, the finite block has distributed mass, and therefore, the block also develops a rotational response.

\begin{equation}
\label{eq:sliding}
m \ddot{u}_x(t) + (c+\mu_s m g) \, \text{sgn} \left[\dot{u}_x(t) \right] = -m \ddot{u}_{g,x}(t)
\end{equation}

The time history response of a block of 100~mm$\times$100~mm in size subjected to a Ricker pulse of frequency $f=2\pi$ rad/s and peak acceleration of $a_p = g$ is shown in Figure \ref{fig:block_dynamic_slide}. The mesh size was $10 \times 10$, therefore a total of 11 of the unit processes were used with an equal spacing of 10 mm. The parameters of the numerical model were $E$, $\nu$, $t$, $K_t$, $K_n$, $\mu$, $c$, and mass density $\rho$ of 10 GPa, $0.3$, 1 mm, $1e^5$ N/mm, $1e^8$ N/mm, 0.4, 1$e^3$ N, and 0.01 kg/mm$^3$, respectively. 

\begin{figure*}[!htb]
\centering
    \begin{subfigure}{.45\textwidth}
    \centering\includegraphics[width=\textwidth]{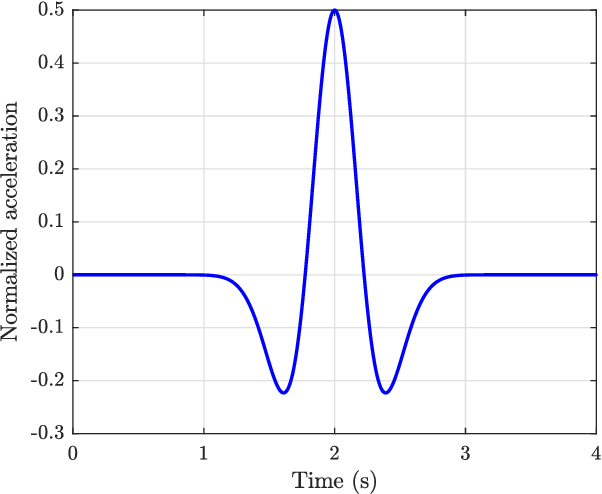}
    \caption{Ricker pulse}
    \label{subfig:sliding_dynamic_ricker}
    \end{subfigure}~
    \begin{subfigure}{.45\textwidth}
    \centering\includegraphics[width=\textwidth]{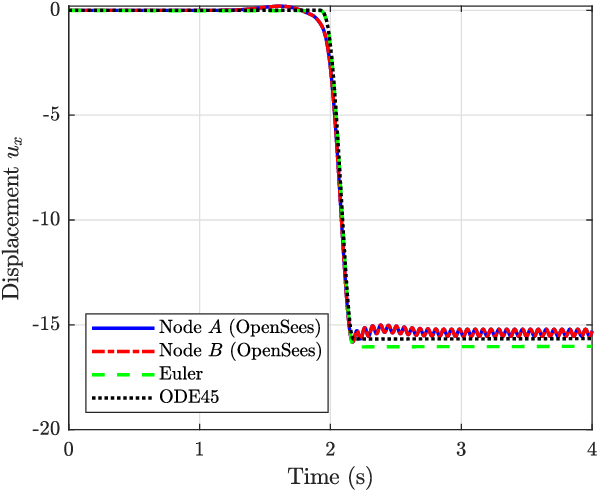}
    \caption{Horizontal displacement}
    \label{subfig:sliding_box_dynamic_dispx}
    \end{subfigure}
    
    \begin{subfigure}{.45\textwidth}
    \centering\includegraphics[width=\textwidth]{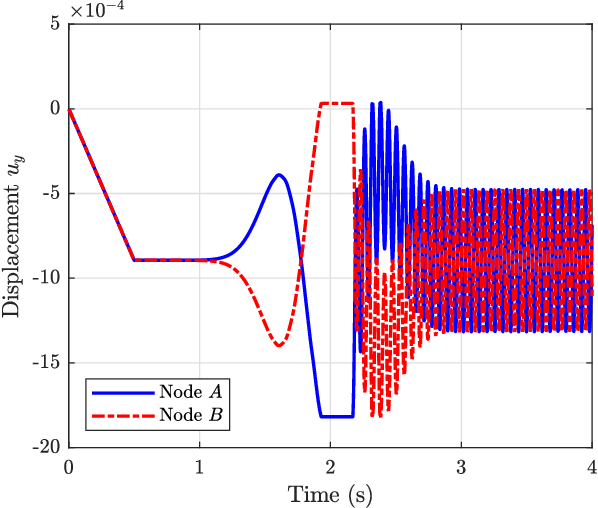}
    \caption{Vertical displacement}
    \label{subfig:sliding_box_dynamic_dispy}
    \end{subfigure}
    \begin{subfigure}{.45\textwidth}
    \centering\includegraphics[width=\textwidth]{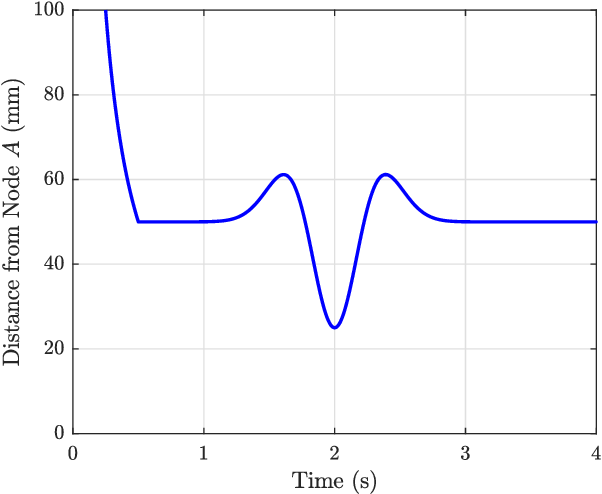}
    \caption{Location of resultant of normal forces}
    \label{subfig:sliding_box_dynamic_resultant_loc}
    \end{subfigure}
    
\caption{Sliding and tipping dynamic case time histories}
\label{fig:sliding_dynamic}
\end{figure*}

The sliding responses of the bottom two corners $A$ and $B$ from OpenSees are shown in Figure \ref{subfig:sliding_box_dynamic_dispx} with numerical solutions by solving Equation \ref{eq:sliding} using Euler method and ode45 in Matlab. The single-degree-of-freedom numerical solutions do not agree exactly with the OpenSees solution due to the rotation of the block that also occurs. The vertical deformation response for the corners are shown in Figure \ref{subfig:sliding_box_dynamic_dispy} where the first 0.5 s represents the application of gravity. Figure \ref{subfig:sliding_box_dynamic_resultant_loc} shows the location of the resultant force during the excitation. For this specific pulse the block response is primarily sliding; however, it is evident that the resultant of the normal force is not always centered on the block during the excitation. 

% The displacements are normalized by $R = \sqrt{h^2 + b^2}$, where $h$ and $b$ are block height and width respectively (do we need to normalize?)
%- slipping initiation 
%- example time histories with mexican hat and ground motion
%- slipping spectrum under symmetric or asymmetric Ricker

% sliding schematic

% need to insert results
% still need some description of what plots show
% need to emphasize that ode45 solution is to a single degree of freedom problem, so the numerical will be different

% ----------------------------------------------------------------------------------------------------------------------------------------
\subsection{Nonlinear Dynamic Rocking}
\begin{figure*}[!htb]
\centering\includegraphics[height=3in]{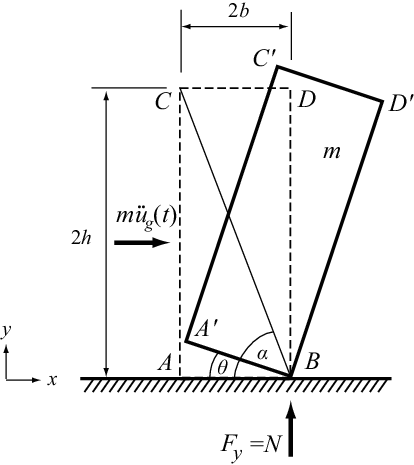}
\caption{Dynamic block rocking schematic}
\label{fig:block_dynamic_rock}
\end{figure*}

Another well-studied problem is that of planar, homogeneous, slender, rigid blocks with translational mass and rotational inertia that uplift and rock under lateral acceleration. The schematic for the rocking block is shown in Figure~\ref{fig:block_dynamic_rock}, where the block has a size defined by $R^2 = h^2 + b^2$ and slenderness $\tan \alpha = b/h$. Once the lateral ground acceleration $\ddot{u}_g$ exceeds $g \tan \alpha$, the block begins to rock/pivot about points $A$ and $B$ in the figure. The equation of motion was previously established for this case in \cite{vassiliou2014dynamic} and is shown in Equation~\eqref{eq:rocking}. The quantities $I_B$, $m$, and $\theta(t)$ are the rotational inertia, translational mass, and rotational angle that correspond to the pivot point $B$ in the figure. 
\begin{equation}
\label{eq:rocking}
I_B \ddot{\theta}(t) + m g R \sin \left[\alpha \text{sgn} \theta(t)-\theta(t) \right] = -m \ddot{u}_{g,x}(t) R \cos \left[\alpha \text{sgn} \theta(t)-\theta(t) \right]
\end{equation}

In the case of the rigid rocking block on rigid foundation, there is no loss of energy from hysteresis. However, energy is lost at each subsequent impact and has been described by others using a coefficient of restitution $r$ that relates the loss of angular velocity immediately before ($\dot{\theta}_1$) and after ($\dot{\theta}_2$) impact. The expression in Equation~\ref{eq:restitution} was solved numerically in tandem with Equation~\ref{eq:rocking} using ode45 in Matlab to verify the behavior of the models. 
\begin{equation}
\label{eq:restitution}
r = \frac{ \dot{\theta}^2_2 } { \dot{\theta}^2_1 } = \left[ 1- \frac{3}{2} \sin^2 \alpha \right]^2
\end{equation}

% rocking schematic

Two different block slenderness ratios were used and subjected to the same symmetric Ricker pulse excitation with $a_p = 3 g \tan \alpha$ and $\omega_p = 2 \pi$ rad/s as were presented in \cite{vassiliou2014dynamic}. The two blocks were defined by $h$ = 5 m and $\tan \alpha$ = 0.1 and 0.2. The mesh size were $10 \times 30$ for both cases, therefore a total of 11 of the unit processes were used. The finite element model was created using planar continuum shell elements with large displacement formulation discretized equally with 10 elements in the $x$ direction and 30 elements in the $y$ direction. The foundation was pinned (single-point constraints). The modulus of elasticity of the block was set at 2.0$e^9$ kPa, Poisson's ratio $\nu$ = 0, and mass density of 7800 kg/m$^3$.

The frictional interface elements were assigned stiff properties: $K_n = 1.0e^{10}$ kN/m, $K_t = 1.0e^{8}$ kN/m, $c = 1.0e^{7}$ kN, and $\mu_s$ = 0.5. The same discretization of the interface elements was used as the location of the block nodes. A $1 g$ constant vertical acceleration was imposed on the model; however, the gravity load was gradually applied over a period of 2 s to avoid oscillations due to the transient analysis. The gravity acceleration is necessary to properly account for the inertia of the block. A mesh refinement study was performed to ensure that the block discretization captured the correct rotational inertia of the block. 

The analysis was performed with a time step of $5.0e^{-4}$ s using the Hilber-Hughes-Taylor (HHT) integrator with a constant of $\alpha_{OS}$ = 0.95, as recommended in \cite{vassiliou2014dynamic} to remove the high frequency vibration generated at impact. 

%- rocking initiation 
%- example time histories with mexican hat and ground motion
%- rocking spectrum under symmetric or asymmetric Ricker, don't think there is any way to do this accurately or within reasonable time

% rocking results
\begin{figure*}[!htb]
\centering
\begin{subfigure}{0.5\textwidth}
\centering\includegraphics[width=\textwidth]{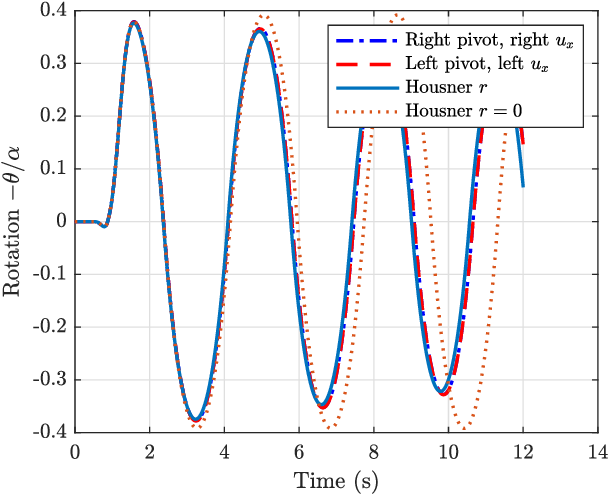}
\caption{Slenderness $\tan \alpha$ = 0.1}
\label{fig:rocking_dynamic_01}
\end{subfigure}~
\begin{subfigure}{0.5\textwidth}
\centering\includegraphics[width=\textwidth]{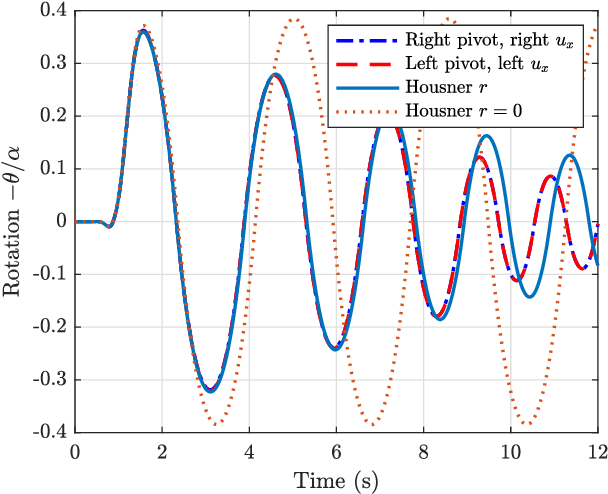}
\caption{Slenderness $\tan \alpha$ = 0.2}
\label{fig:rocking_dynamic_02}
\end{subfigure}
\caption{Rocking block rotation histories under Ricker pulse excitation. Analytical solution shown for cases with and without coefficient of restitution.}
\label{fig:rocking_dynamic}
\end{figure*}

% Case studies using dry friction solved using proposed method
% Case Studies
% status: incomplete

\section{Case Studies using Dry Friction}
Three case study problems are presented here to showcase the dry friction implementation. 

\subsection{Rabinowicz Test Case}
\begin{figure*}[!htb]
    \centering
    \begin{subfigure}{.45\textwidth}
        \centering\includegraphics[width=0.6\textwidth]{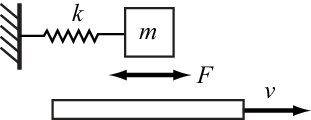}
        \caption{Rabinowicz test case model}
        \label{fig:rabinowicz_model}
    \end{subfigure}~
    \begin{subfigure}{.45\textwidth}
        \centering\includegraphics[width=\textwidth]{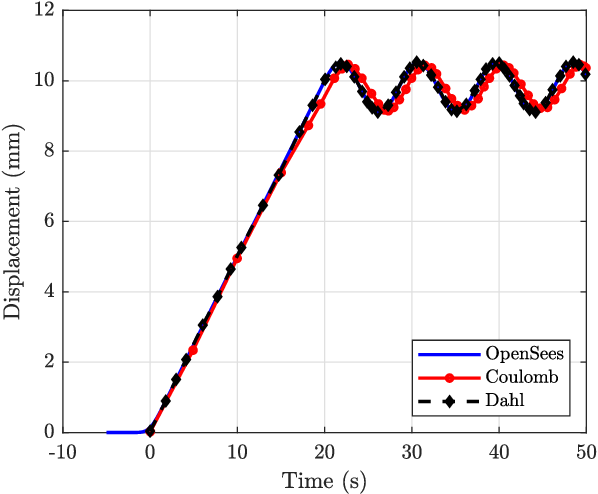}
        \caption{Displacement}
        \label{fig:rabinowicz_disp}
    \end{subfigure}
    
    \begin{subfigure}{.45\textwidth}
        \centering\includegraphics[width=\textwidth]{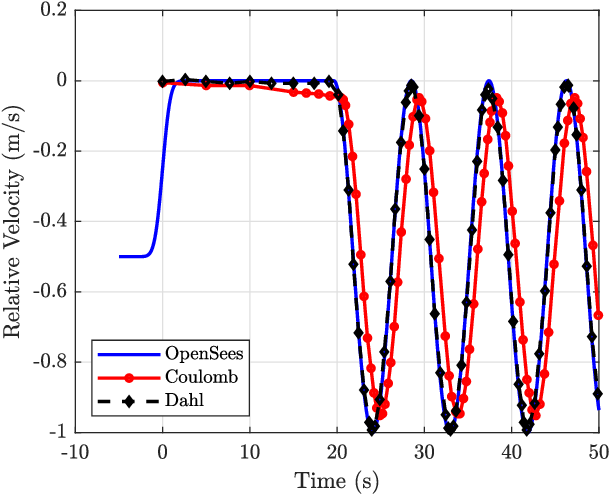}
        \caption{Velocity}
        \label{fig:rabinowicz_vel}
    \end{subfigure}~
    \begin{subfigure}{.45\textwidth}
        \centering\includegraphics[width=\textwidth]{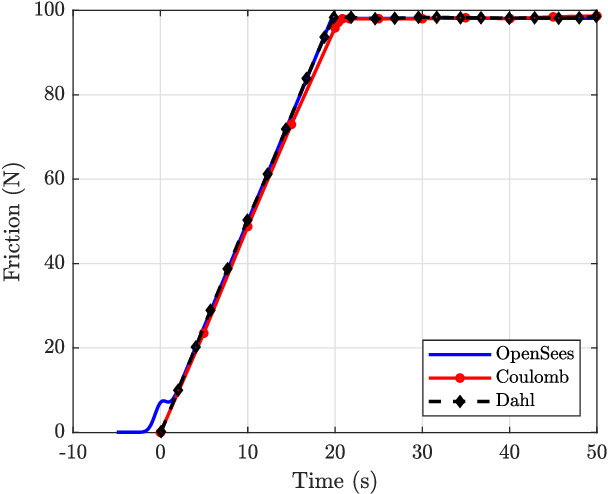}
        \caption{Friction}
        \label{fig:rabinowicz_friction}
    \end{subfigure}

    \caption{Rabinowicz case time histories}
    \label{fig:rabinowicz}
\end{figure*}

 A benchmark case for evaluating friction models is the Rabinowicz test as shown in Figure \ref{fig:rabinowicz_model} where a mass $m$ is connected by a spring of stiffness $k$. The mass rests over a slab which is moving by a constant velocity $v$. At the beginning, the block is dragged by the slab until the spring force is equal to the friction force at the interface. As soon as the interface friction reaches its maximum value, the block is pulled by the spring and starts oscillating back and forth. After some time, the block stops oscillating and the static friction takes over which causes it to move along the slab again and the process repeats.

% rabinowicz model and plot

The Rabinowicz test case was simulated by \citet{pennestri2016review} for the parameters $m=20$ kg, $k=10$ N/m, $v=0.5$ m/s, a Coulomb static friction coefficient $\mu_s=0.6$, and a Coulomb dynamic friction coefficient $\mu_d=0.5$. The same case was implemented in OpenSees using the Newmark integration scheme. Instead of applying velocity directly, a time-varying displacement was imposed on the mass to obtain the desired velocity. The velocity was applied as a gradual ramp to prevent any oscillation in the response.

% let's read in the paper the descriptions for Dahl and Coulomb and why they are so named
The responses obtained from the proposed friction element model is shown in Figure \ref{fig:rabinowicz} with the responses from \citet{pennestri2016review} for Coulomb and Dahl friction model - the agreement with the proposed model being with Dahl in this case (as the Coulomb results did not have constant velocity as reported). The Figure \ref{fig:rabinowicz_disp} shows the displacement of the block versus time, Figure \ref{fig:rabinowicz_vel} shows the relative velocity between the block and the slab versus time, and finally Figure \ref{fig:rabinowicz_friction} shows the friction force developed at the interface versus time.

\subsection{Steel bolted joint}

\begin{figure*}[!htb]
    %\centering
    %\begin{subfigure}{.52\textwidth}
        \centering\includegraphics[width=0.3\textwidth]{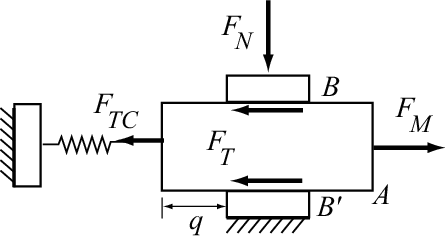}
        %\caption{Model}
        %\label{fig:bolted_joint_model}
    %\end{subfigure}~
    %
    %\begin{subfigure}{.45\textwidth}
    %    \centering\includegraphics[width=\textwidth]{figures/rabinowicz/rabinowicz_disp.eps}
    %    \caption{Displacement}
    %    \label{fig:rabinowicz_disp}
    %\end{subfigure}
    
    \caption{Bolted joint model}
    \label{fig:bolted_joint}
\end{figure*}

Figure \ref{fig:bolted_joint} shows the schematic of an experimental study conducted by \cite{ferrero2004analysis}. A block $A$ is under normal force $F_N$ on top and bottom from a bolted joint $BB^\prime$. It is also attached to a load sensor at left which can record the force $F_{TC}$ developed at the sensor. An imposed external load $F_M$ is applied on the block from the right that causes it to slide a distance $q$. The frictional forces developed at the two interfaces can be computed from $F_T = (F_M-F_{TC})/2$.

% bolted joint model and plot

\begin{figure*}
\centering
    \begin{subfigure}{.50\textwidth}
    \centering\includegraphics[width=\textwidth]{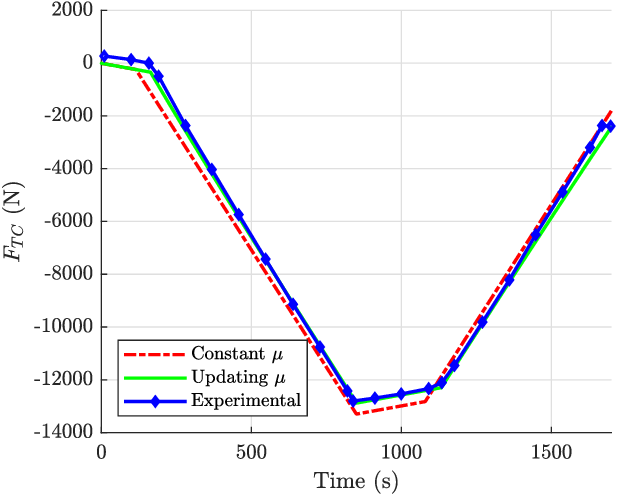}
    \caption{Time vs. $F_{TC}$, $F_{N} = 2$ kN}
    %\label{fig:block_bottom_1}
    \end{subfigure}~~
    \begin{subfigure}{.50\textwidth}
    \centering\includegraphics[width=\textwidth]{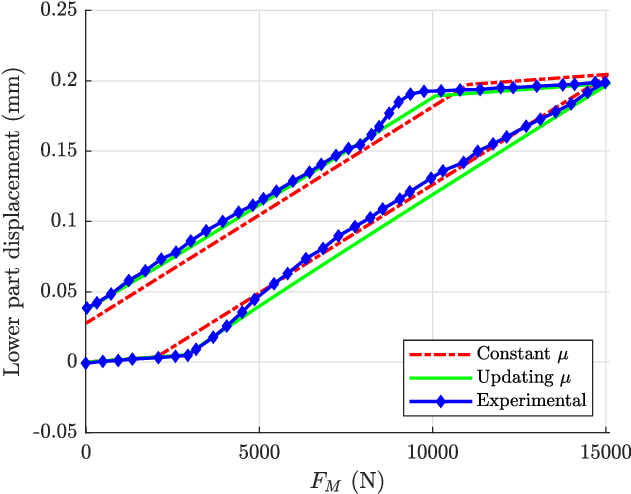}
    \caption{$F_{M}$ vs. displacement, $F_{N} = 2$ kN}
    %\label{fig:block_bottom_2}
    \end{subfigure}
    
    \begin{subfigure}{.50\textwidth}
    \centering\includegraphics[width=\textwidth]{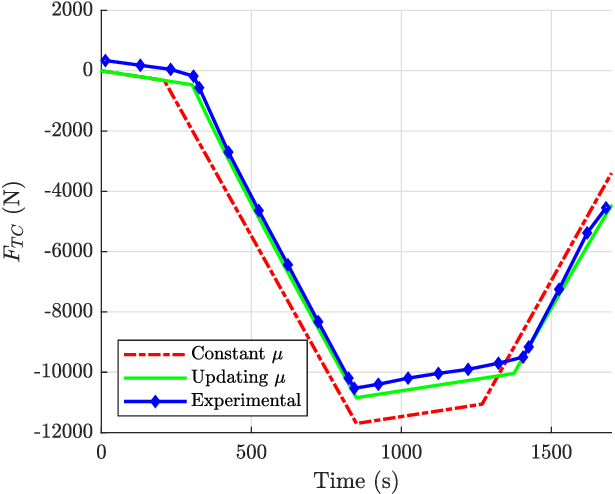}
    \caption{Time vs. $F_{TC}$, $F_{N} = 4$ kN}
    %\label{fig:block_bottom_1}
    \end{subfigure}~~
    \begin{subfigure}{.50\textwidth}
    \centering\includegraphics[width=\textwidth]{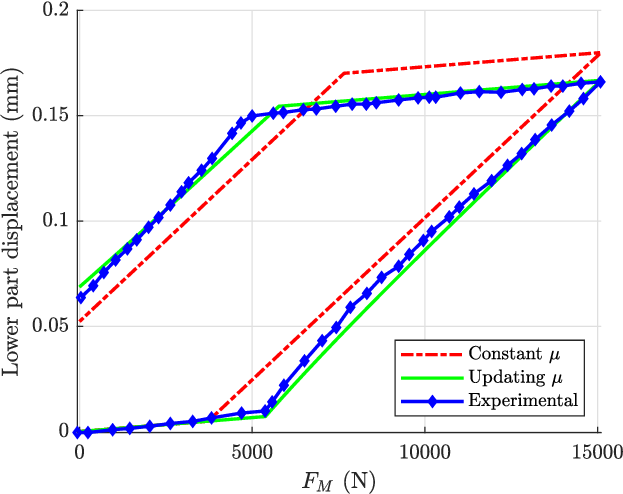}
    \caption{$F_{M}$ vs. displacement, $F_{N} = 4$ kN}
    %\label{fig:block_bottom_2}
    \end{subfigure}
    
    \begin{subfigure}{.50\textwidth}
    \centering\includegraphics[width=\textwidth]{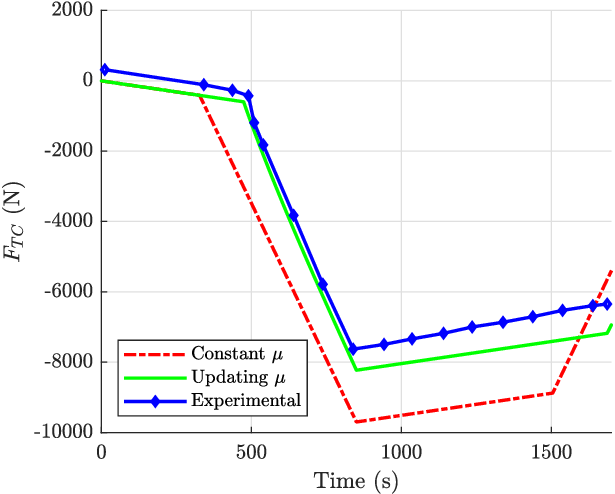}
    \caption{Time vs. $F_{TC}$, $F_{N} = 6$ kN}
    %\label{fig:block_bottom_1}
    \end{subfigure}~~
    \begin{subfigure}{.50\textwidth}
    \centering\includegraphics[width=\textwidth]{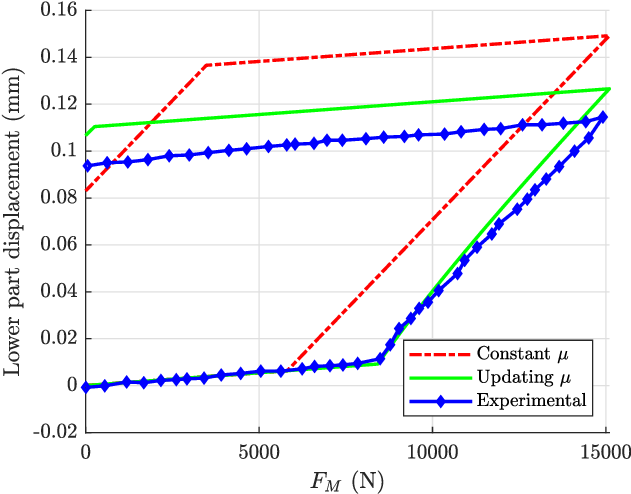}
    \caption{$F_{M}$ vs. displacement, $F_{N} = 6$ kN}
    %\label{fig:block_bottom_2}
    \end{subfigure}
    
\caption{Bolted joint time histories}
\label{fig:bolted_joint_plot}
\end{figure*}

Figure \ref{fig:bolted_joint_plot} compares the numerical results with the experimental results obtained from \cite{ferrero2004analysis} for three different cases of normal force $F_N$. The numerical results are presented for two modeling cases: one where frictional coefficient $\mu$ remains constant, and an updated $\mu$ that depends on the sliding distance. The updated frictional coefficient is computed using the relationship $\mu = \mu_d + (\mu_c - \mu_d) e^{-Cq}$, where $\mu_d$ and $\mu_c$ are the smallest and largest value of $\mu$ respectively, $C$ is a numerical coefficient and $q$ is the sliding distance (with values of the coefficients calibrated by the original authors of the experiment). The results for the updated $\mu$ case closely follow the experimental data for $F_N$ = 2 kN, 4 kN cases; however, they deviate slightly for the 6 kN case.

\subsection{Cantilevers in Contact}
\begin{figure*}
    \centering
        \begin{subfigure}{.30\textwidth}
            \centering\includegraphics[width=\textwidth]{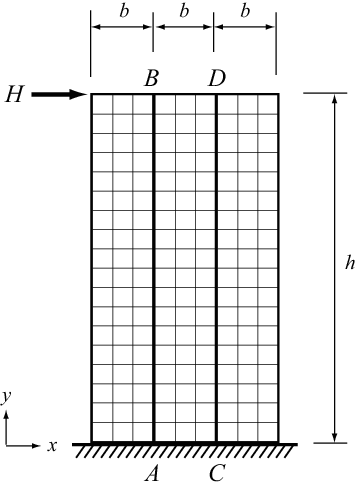}
            \caption{Three cantilever model}
            \label{fig:cantilever_schematic}
        \end{subfigure}~
        \begin{subfigure}{.33\textwidth}
            \centering\includegraphics[width=0.80\textwidth]{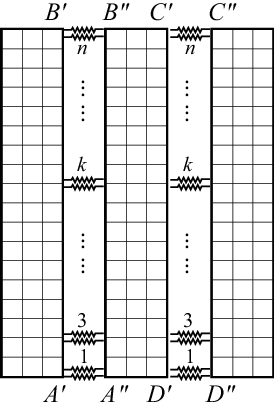}
            \caption{Frictional elements}
            \label{fig:cantilever_fbd}
        \end{subfigure}~
        \begin{subfigure}{.30\textwidth}
            \centering\includegraphics[width=0.5\textwidth]{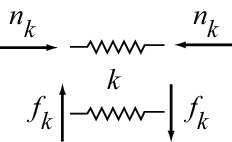}
            \caption{Element forces at $k$-th element in interface $AB$}
            \label{fig:cantilever_element}
        \end{subfigure}
    \caption{Three cantilever model}
    \label{fig:three_cantilever}
\end{figure*}

Three independent cantilever columns modeled with planar continuum isoparametric elements having two contact surfaces $AB$ and $CD$ are analyzed, as shown in Figure \ref{fig:cantilever_schematic}. The frictional elements $k$ are placed at each node along the contact surfaces on both interfaces $AB$ and $CD$. One component of $k$ corresponds to normal forces $n_k$ and the other component takes care of the frictional forces $f_k$ at the same node. Each cantilever has width $b=200$ mm, height $h=6000$ mm and out of plane thickness of 1000 mm. The modulus of elasticity is $E=24e^6$ N/mm$^2$ and Poisson's ratio is $\nu=0.01$. A reference lateral point load of $H=1e^7$ N was applied at the top of the left cantilever. For large $c$ and $\mu$, the three cantilevers act as a single composite cantilever with strain compatibility across the interfaces. For very small $c$ and $\mu$, the three cantilevers bend separately and the tip deflection is significantly larger than the single cantilever case.

% three cantilever figures

\begin{figure*}
    \centering
    \begin{subfigure}{0.5\textwidth}
        \centering\includegraphics[width=\textwidth]{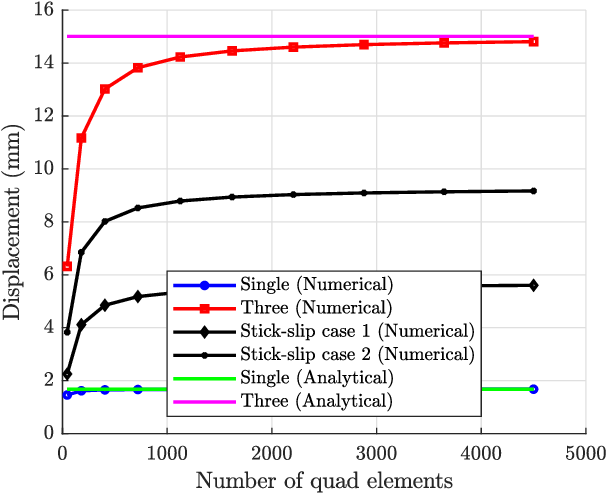}
        \caption{Convergence analysis}
        \label{fig:three_cantilever_convergence}
    \end{subfigure}~
    \begin{subfigure}{0.5\textwidth}
        \centering\includegraphics[width=\textwidth]{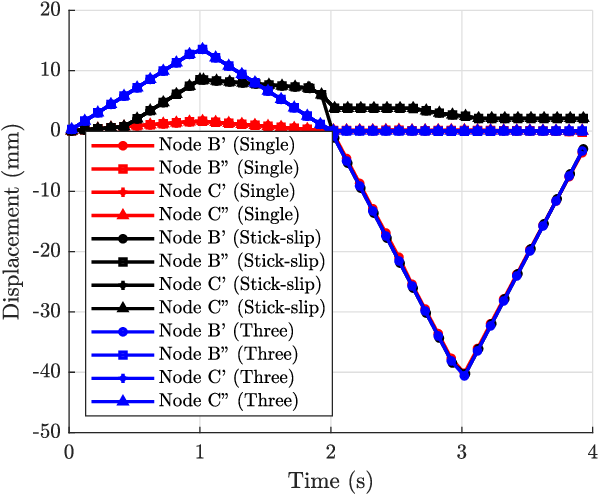}
        \caption{Time history}
        \label{fig:three_cantilever_timehistory}
    \end{subfigure}~
    \caption{Three cantilevers response}
    \label{fig:three_cantilever_response}
\end{figure*}

Figure \ref{fig:three_cantilever_convergence} shows the convergence analysis for different mesh discretizations. The number of vertical subdivisions for each cantilever was six times its horizontal subdivision. The horizontal subdivisions were increased by 3 starting from 3 to 30 (therefore a total of 11 models for each case). For example, the first model had $3 \times 18$ mesh size (or 54 quad elements); therefore a total of 19 unit processes were used for each interface.
%The second model had $6 \times 36$ mesh size (or 216 quad elements), a total of 37 unit processes were used for each interface and so on. 
The final model had $30 \times 180$ mesh size (or 5400 quad elements) and a total of 181 unit processes for each interface. 

The reference load was monotonically increased to the maximum value listed above. As the number of quadrilateral (Q4) elements was increased, the displacements of the single and three cantilever cases approached the closed-form solutions of 1.677 mm and 15 mm respectively. Two stick-slip cases were implemented with parameter values of $\mu = 0.4$ and $c = 5 e^6$ for stick-slip case 1, and $\mu = 0.4$ and $c = 3 e^6$ for stick-slip case 2. In the stick-slip cases, the system initially behaves as a single (composite) cantilever. However, after slipping initiates and progresses along the interfaces, it behaves more similar to the three cantilever system, but with residual frictional resistance at the locations of contact. 

A second load pattern was investigated that comprised of a single sawtooth cycle with a period of 4 s. Figure \ref{fig:three_cantilever_timehistory} shows the time history of the tip displacements of the interface nodes. Only one of the stick-slip cases was retained for simplicity (corresponds to stick-slip 2 above for mesh subdivision $12 \times 72$ for each cantilever). For the single and three cantilever cases, all three columns moved together in the first half cycle from 0 to 2 s. In the second half cycle from 2 s to 4 s, only the left-most cantilever moved while the other two remained in their original position. For the stick-slip case, the three cantilevers started to deform as a single composite column, with the same initial stiffness. Once slip occurred, the stiffness followed that of the three cantilever system. Unloading again mimicked the single cantilever case until slipping occurred again. When the load reversed, only the left-most cantilever moved but the other two had some residual displacements due to the additional resistance stored at the interface locations (slipping was not symmetric). 
%Eventually the right two cantilevers slip again and return to zero displacement.

%\subsection{Extrusion}
%Lack of fit/extrusion case.
%NYI

%\subsection{Beam on Elastic Frictional Foundation}
%Traditional BEF theory but adding a foundation with horizontal frictional term.
%NYI

%\subsection{Friction Pendulum}
%double or triple concave bearings with known mathematical and experimental solutions
%NYI

%\subsection{Inverted pendulum on sliding block}
%This is based on a paper with presumeably an analytical and numerical solution. Has more to do with inverted pendulum than sliding block. 
% NYI

% Wrap-up and review
\section{Conclusion}
The paper demonstrated the implementation of a node-to-node Coulomb friction interface using only existing finite elements, specifically a two-noded element with either an elastic or elasto-plastic force-deformation (constitutive) relation. The contact-friction implementation and application to a series of validation problems was achieved using run-time parameter updates to the stiffness or strength of the elastic or elasto-plastic constitutive models in OpenSees, respectively. The implication of such a run-time parameter-driven system is that it removes the need for finite element software to code physics or material-specific materials, elements, or algorithms in similar deployment scenarios to contact-friction. There is value added in solving classical problem using existing finite elements, because of the barriers of commercial black-box systems or theory, coding, and debugging in open-source systems. 

The contact-friction parameter implementation was demonstrated for simple two-noded static and dynamic problems and shown to achieve the closed-form solutions. The unit process was then replicated to a finite length interface for a rigid block on rigid foundation for sliding, slipping, tipping, and rocking analyses. Finally, several case studies were implemented, including comparison with experimental results and an interface between deformable cantilevers in bending. The code can be packaged in a class or function to replicate in more complex problems like bridge abutments. The current examples are limited to two-dimensional node-to-node contact. The relative deformations between the nodes can be large, and the outward normal can be updated to account for flexible contact surfaces. However, the model has the same limitations as any another other node-to-node model when sliding deformations are large. 

The contribution to dry friction problems in structural engineering is that extensions can be readily made to non-classical friction cases. This is a natural extension of the parameter implementation, as the frictional component of response can be specified in terms of any uniaxial constitutive model. Therefore, viscoplastic models such as Dahl or LuGre can be implemented using parallel arrangements of viscous and elasto-plastic constitutive models.

%\appendix

%\section{Sample Appendix Section}
%Lorem ipsum dolor sit amet, consectetur adipiscing elit, sed do eiusmod tempor section \ref{sec:sample1} incididunt ut labore et dolore magna aliqua. Ut enim ad minim veniam, quis nostrud exercitation ullamco laboris nisi ut aliquip ex ea commodo consequat. Duis aute irure dolor in reprehenderit in voluptate velit esse cillum dolore eu fugiat nulla pariatur. Excepteur sint occaecat cupidatat non proident, sunt in culpa qui officia deserunt mollit anim id est laborum.

\bibliographystyle{elsarticle-num-names} 
\bibliography{ref}





% \end{thebibliography}
\end{document}